\documentclass[twocolumn,prl,superscriptaddress]{revtex4}
\usepackage{amsmath}
\usepackage{graphicx}
\usepackage{bm}
\usepackage{fancyhdr}
\usepackage{amsmath}
\usepackage{graphicx}
\usepackage{amsfonts}
\usepackage{amsbsy}
\usepackage{amssymb}
\usepackage[mathscr]{eucal}
\usepackage{color}

\newcommand{\beq}{\begin{equation}}
\newcommand{\eeq}{\end{equation}}
\newcommand{\ba}{\begin{array}}
\newcommand{\ea}{\end{array}}
\newcommand{\bea}{\begin{eqnarray}}
\newcommand{\eea}{\end{eqnarray}}
\newcommand{\bseq}{\begin{subequations}}
\newcommand{\eseq}{\end{subequations}}

\begin{document}

\title{Determination of weak values of quantum operators using only
strong measurements}
\author{Eliahu Cohen}
\affiliation{Physics Department, Centre for Research in Photonics, University of Ottawa,                                                  Advanced Research Complex, 25 Templeton, Ottawa ON Canada, K1N 6N5}
\affiliation{Faculty of Engineering and the Institute of Nanotechnology and Advanced
Materials, Bar Ilan University, Ramat Gan 5290002, Israel}

\author{Eli Pollak}
\affiliation{Chemical and Biological Physics Department, Weizmann Institute of
Science, 76100, Rehovot, Israel}

\begin{abstract}
Weak values have been shown to be helpful especially when
considering them as the outcomes of weak measurements. In this paper we
show that in principle, the real and imaginary parts of the weak value of
any operator may be elucidated from expectation values of suitably defined
density, flux and hermitian commutator operators. Expectation values are the outcomes of strong (projective)
measurements implying that weak values are general properties of operators
in association with pre- and post-selection and they {  need not} be
preferentially associated with weak measurements. They should be considered
as an important measurable property which provides added information as
compared with the ``standard'' diagonal expectation value of an operator. {  As a first specific example} we consider the determination of the real and imaginary
parts of the weak value of the momentum operator employing projective time of flight experiments. {  Then the results are analyzed from the point of view of Bohmian mechanics. Finally we consider } recent neutron interferometry experiments used to
determine the weak values of the neutron spin.
\end{abstract}

\maketitle

\section{I. Introduction}
Weak values naturally appear as a result of weak measurement
when one considers pre- and post-selected systems \cite{aharonov88}. For an
initial pre-selected state $|\Psi\rangle$ at time $t_i$, evolved to an
intermediate time $t$, and a final post-selected state $|\Phi\rangle$ at
time $t_f$ (which may or may not be the same as $t$), the weak value of the
operator ${\hat A}$ is defined as \cite{aharonov88,aharonov14}
\begin{equation}
\langle \hat A \rangle_w (t) = \frac{\langle \Phi(t) | \hat A |
\Psi(t)\rangle}{\langle \Phi(t) | \Psi(t)\rangle} .  \label{1}
\end{equation}
The relation between weak measurement and weak values was
derived by using a linear approximation to unitary time evolution when the
coupling of the measurement apparatus to the pre- and post-selected system
is weak enough \cite{aharonov88,kofman12,dressel14}. This is
the source of the nomenclature of ``weak values''. It is therefore not surprising
that subsequently, weak values have been commonly measured using weak
measurements, see e.g. \cite{ritchie91,pryde05,steinberg10}.

The introduction of the weak value concept has had a profound impact on our
understanding of quantum mechanics. It led to the development of new
phenomena such as quantum random walks \cite{aharonov93} and
superoscillations \cite{berry94a,berry94b}. It has influenced recent
theoretical \cite{aharonov05,lund10,matzkin12,dressel14a,cohen17a,georgiev18}
and experimental \cite{goggin11,kocsis11,danan13,piacentini16} studies of
quantum foundations. The weak value has been an important tool in the
development of precision measurements \cite%
{hosten08,dixon09,starling10,xu13,pang15}, as well as state \cite%
{lundeen11,thek16} and process \cite{reznik13,kim18} tomography.

Yet the concepts of a weak value, and the related weak
measurement are controversial to this very date \cite%
{jordan14,romito16,vaidman17}. It has been claimed that the definition of a
weak value is a mere generalization of the notion of an expectation value to
the case of differing pre- and post-selected states but that it does not
provide much insight into physical reality, e.g. \cite%
{parrott09,sokolovski13,svensson13}. Others note that weak values and weak
measurements have provided and continue to provide interesting physical
insights \cite{kofman12,yokota09,lundeen12,dressel15,white16,cohen17b},
going beyond the notion of a generalized expectation value \cite{vaidman17b}. Yet it is still claimed that since the weak value is inevitably linked to
a weak measurement involving a ``meter'' it depends not only on the measured
quantum system but also on the measuring meter \cite{svensson14}.

There have been in recent years a growing number of works \cite%
{johansen07,hiley12,berry13,vallone16,zhang16} which consider inferring weak
values using (strong) projective measurements \cite{denkmayr17,sponar18}, yet not with full generality. In this paper we prove via a new and general protocol that both the real and
imaginary parts of weak values can be obtained in principle through strong projective
measurements. We thereby disconnect the concept of weak value from the
concept of weak measurement, enhancing the validity and applicability of the former.

The paper is organized as follows. We present in Sec. II the general formalism for inferring the weak value of any operator. Then, in Secs. III-VI we consider in detail the special case of obtaining the weak value of the momentum from projective measurements of the density and the flux operators. We utilize in our analysis the concept of transition path time distribution \cite{Pollak2017c,petersen17}, as well as time of flight experiments. As a first application of our results we revisit in Sec. VII the role of weak values in Bohmian mechanics. As a second application of our formalism we analyze in Sec. VIII recent experiments employing neutron
interferometry \cite{denkmayr17,sponar18}. We end in Sec. IX with a discussion on the
implications of these results on the general weak value formulation of quantum mechanics.

We stress that the aim of this work is not to dismiss the physical origin of weak values as being associated with a shift of a pointer weakly coupled to a pre- and post-selected system. We find this traditional understanding interesting and profound. We rather wish to broaden the meaning of weak values and extend the measurement techniques commonly used for inferring them.




\section{II. Inferring weak values from strong measurements}

Consider the operator $\hat{A}$ and its weak value as defined in Eq. \ref{1}
for the pre-selected state $|\Psi \rangle $ at time $t_{i}$ and a
post-selected state $|\Phi \rangle $ at time $t$. The hermitian density
operator related to the post-selected state is by definition%
\begin{equation}
\hat{D}\left( \Phi \right) =\left\vert \Phi \rangle \langle \Phi \right\vert
.  \label{2}
\end{equation}%
We then define a generalized hermitian ``flux'' operator
associated with the post-selected state and the operator $\hat{A}$ as the
(hermitian) anti-commutator of the operator $\hat{A}$ and the density
operator
\begin{equation}
\hat{F}\left( \Phi \right) =\frac{1}{2}\left\{ \hat{A},\hat{D}\left( \Phi
\right) \right\} \equiv \frac{1}{2}\left( \hat{A}\hat{D}\left( \Phi \right) +%
\hat{D}\left( \Phi \right) \hat{A}^{\dag }\right)  \label{3}
\end{equation}%
We also define the hermitian commutator operator
\begin{equation}
\hat{C}\left( \Phi \right) =\frac{1}{2}\left[ i\hat{A},\hat{D}\left( \Phi
\right) \right] \equiv \frac{i}{2}\left( \hat{A}\hat{D}\left( \Phi \right) -%
\hat{D}\left( \Phi \right)\hat{A}^{\dag }\right)  \label{4}
\end{equation}
It is then a matter of straightforward calculation to prove that
\begin{equation}
\frac{\langle \Psi |\hat{F}\left( \Phi \right) |\Psi \rangle }{\langle \Psi |%
\hat{D}\left( \Phi \right) |\Psi \rangle }=\mathrm{Re}\langle \hat{A}{(\Phi
;\Psi )}\rangle _{w}  \label{5}
\end{equation}%
and%
\begin{equation}
\frac{\langle \Psi |\hat{C}\left( \Phi \right) |\Psi \rangle }{\langle \Psi |%
\hat{D}\left( \Phi \right) |\Psi \rangle }=\mathrm{Im}\langle \hat{A}{(\Phi
;\Psi )}\rangle _{w}.  \label{6}
\end{equation}%
We have thus demonstrated in very general terms that the real
and imaginary parts of the weak value of an operator can be obtained
through at most three strong projective measurements. The practical question of how one
implements them for the relevant operators depends on
the identity of the operator $\hat{A}$, as well as the pre- and
post-selected states and is not necessarily trivial. However, any weak value
associated with the operator $\hat{A}$ may be inferred \textit{in principle}
from strong measurements. We will now consider the specific
example of the weak value of the momentum operator, this example will also
explain why we relate to the anti-commutator (Eq. \ref{3}) as a generalized
``flux'' operator.

\section{III. Momentum weak values through strong measurements}

We limit ourselves to a one dimensional particle, with mass $M
$, whose time evolution is determined by the Hamiltonian
\begin{equation}
{\hat H}=\frac{{\hat p}^2}{2M}+V({\hat q}),  \label{7}
\end{equation}
where 
${\hat q}$ and ${\hat p}$ are the coordinate and momentum operators,
respectively. The density and flux hermitian operators at the point $x$ are
defined as usual as:
\begin{equation}
\hat{D}\left( x\right) =\delta \left( \hat{q}-x\right)  \label{8}
\end{equation}
\begin{equation}
\hat{F}\left( x\right) =\frac{1}{2M}\left[ \hat{p}\delta\left( \hat{q}%
-x\right) +\delta\left( \hat{q}-x\right)\hat{p}\right].  \label{9}
\end{equation}%
Note the parallelism between these standard definitions and their generalization as expressed in Eqs. \ref{2} and \ref{3}.

We are interested in the weak value of the momentum at a post-selected point
$x$ using the pre-selected (normalized) state $| \Psi \rangle $:
\begin{equation}
\langle {\hat p(x;\Psi)}\rangle_w=\frac{ \left\langle x\left\vert {\hat p}
\right\vert \Psi \right\rangle }{\left\langle x|\Psi \right\rangle }.
\label{10}
\end{equation}
It is a matter of straightforward calculation, using Eqs. \ref%
{2}-\ref{6} to derive the following three identities:
\begin{eqnarray}
\noindent \langle \Psi |\hat{D}\left( x\right) |\Psi \rangle &=&\left\vert
\left\langle x|\Psi \right\rangle \right\vert ^{2}  \label{11} \\
\frac{\langle \Psi |\hat{F}\left( x\right) |\Psi \rangle }{\langle \Psi |%
\hat{D}\left( x\right) |\Psi \rangle }&=&\frac{\mathrm{Re}\langle {\hat{p}%
(x;\Psi )}\rangle _{w}}{M}  \label{12}  \\
\frac{1}{2}\frac{\langle \Psi |\left[ i{\hat{p},}\hat{D}\left( x\right) %
\right] |\Psi \rangle }{\langle \Psi |\hat{D}\left( x\right) |\Psi \rangle }
&=&\mathrm{Im}\langle {\hat{p}(x;\Psi )}\rangle _{w}. \label{13}
\end{eqnarray}%
This shows explicitly that the real and imaginary parts of the weak value of
the momentum may be determined with only strong measurements.
We shall now demonstrate, using a transition path time distribution approach, how one may
in principle measure the flux and hermitian commutator operators using strong measurements.

\section{IV. A transition path time distribution}

We consider a scattering experiment, such that the potential goes to
constant values as $x\rightarrow \pm\infty$. The particle is prepared
initially at time $t=0$ to be in the state $|\Psi _{0}\rangle $ localized
around an initial position $y$ and (positive) momentum $p_{y}$, say to the left of the
potential. The pre-selected state $|\Psi _{0}\rangle $ may for example be
the coherent state:%
\begin{equation}
\left\langle q|\Psi _{0}\right\rangle =\left( \frac{\Gamma }{\pi }\right)
^{1/4}\exp \left[ -\frac{\Gamma }{2}\left( q-y\right) ^{2}+i\frac{p_{y}}{%
\hbar }\left( q-y\right) \right] .  \label{14}
\end{equation}%
We then post-select a position $x$ to the right of the potential and measure
the time $t$ at which the particle reaches this position. In this scenario,
we set $t_{f}=t$, that is the intermediate time $t$ at which the weak valued
is inferred in Eq. \ref{1} is identical to the final time at which the
post-selection takes place. The probability density $\rho \left( x|t\right) $
for the particle to reach the position $x$ at the time $t$ is%
\begin{equation}
\rho \left( x|t\right) =\left\vert \left\langle x|\Psi _{t}\right\rangle
\right\vert ^{2},  \label{15}
\end{equation}%
where
\begin{equation}
|\Psi _{t}\rangle =\exp \left( -\frac{i}{\hbar }{\hat H}t\right)|\Psi
_{0}\rangle  \label{16}
\end{equation}
is the time evolved pre-selected state. The distribution $\rho$ is
normalized
\begin{equation}
\int_{-\infty }^{\infty }dx\rho \left( x|t\right) =1.  \label{17}
\end{equation}

One may also define the probability density\ $\rho \left( t|x\right) $ for
the distribution of times at which the particle will reach the post-selected
point $x$. It is given by the transition path time distribution \cite{Pollak2017c,petersen17}
\begin{equation}
\rho \left( t|x\right) =\frac{\left\vert \left\langle x|\Psi
_{t}\right\rangle \right\vert ^{2}}{\int_{0}^{\infty }dt\left\vert
\left\langle x|\Psi _{t}\right\rangle \right\vert ^{2}}\equiv \frac{%
\left\vert \left\langle x|\Psi _{t}\right\rangle \right\vert ^{2}}{N\left(
x\right) }  \label{18}
\end{equation}%
and by definition
\begin{equation}
\int_{0}^{\infty }dt\rho \left( t|x\right) =1.  \label{19}
\end{equation}%
$\rho \left( t|x\right) $ is termed the transition path time probability
distribution associated with the pre-selected state $|\Psi _{0}\rangle $ and
the post-selected position $x$. This time distribution is in principle
measurable by sufficient repetition of a single atom time of flight apparatus \cite{Fuhrmanek2010},
that measures the time $t=0$ at which a particle, prepared in the state $%
|\Psi _{0}\rangle $, exits a source \cite{Du2015}, and then the time $t$ at
which it reaches the detector located at $x$.

To measure $\rho \left( t|x\right)$ at any point $x$, one may place a
detector at $x$ and divide a reasonably long time interval $T$ into $N$
equal steps $t_n=n\Delta T$, where $n \in \mathbb{N}$, such that $N\Delta
T=T $. This will enable one to obtain in a coarse grained fashion the
spatial derivative $\frac{\partial \rho \left( t|x\right)}{\partial x}$ of
the transition path time distribution using finite differences $\Delta x$ in
space.

Aharonov \textit{et al.} \cite{aharonov98} have shown that the time of
arrival cannot be measured more accurately than $\Delta t \approx \hbar/E_k$%
, where $E_k$ is the initial kinetic energy of the particle. Current
detectors of massive particles typically have temporal resolution of
picoseconds \cite{ebran13}, so for kinetic energy larger than $10^{-22}$ J,
this temporal resolution can be met. For neutrons, this implies a
non-relativistic velocity of (at least) $v\approx 350$ m/s which is not
extremely high.

\section{V. Inferring the imaginary part of the weak value of the momentum}

Consider then a time of flight measurement of the distribution, once at $%
x-\Delta x/2$ and then at $x+\Delta x/2$.

Noting that the coordinate representation of the momentum operator is such
that
\begin{equation}
\left\langle x\left\vert {\hat{p}}\right\vert \Psi \right\rangle =-i\hbar
\frac{\partial }{\partial x}\left\langle x|\Psi \right\rangle,  \label{20}
\end{equation} one readily finds that:%
\begin{eqnarray}
\frac{\rho \left( t|x+\frac{\Delta x}{2}\right) -\rho \left( t|x-%
\frac{\Delta x}{2}\right) }{\Delta x} \simeq -\frac{%
\partial \ln N\left( x\right) }{\partial x} \notag \\ -
\frac{2}{\hbar }\rho \left( t|x\right) \mathrm{Im}\left[ \frac{\left\langle
x\left\vert p\right\vert \Psi _{t}\right\rangle }{\left\langle x|\Psi
_{t}\right\rangle }\right] +O\left( \Delta x\right) .  \label{21}
\end{eqnarray}
In most scattering cases, if the post-selected position $x$ is sufficiently
far out in the asymptotic region, the normalization $N\left( x\right) $
becomes independent of $x$ \cite{pollak18} so that measuring the transition
path time distribution at the post-selected positions $x-\frac{\Delta x}{2}%
,x $ and $x+\frac{\Delta x}{2}$ allows the direct determination (without
invoking weak measurements) of the imaginary part of the weak value of the
momentum at the position $x$:
\begin{eqnarray}
\mathrm{Im}\left[ \frac{\left\langle x\left\vert p\right\vert \Psi
_{t}\right\rangle }{\left\langle x|\Psi _{t}\right\rangle }\right]\simeq-%
\frac{\hbar}{2}\frac{\partial \ln\rho \left( t|x \right)}{\partial x}=-\frac{%
\hbar}{2}\frac{\partial \ln\rho \left( x|t \right)}{\partial x}  \label{22}
\end{eqnarray}
and this is identical to the result given in Eq. \ref{13}. The time of flight
measurement therefore provides an experimentally implementable protocol for
obtaining the imaginary part of the weak value of the momentum. Even if the
normalization is a function of $x$ it is of course time-independent so that
it just serves as a constant base line which may be subtracted out.

\subsection{Further notes regarding the imaginary part of the momentum weak value}

One of the challenges posed by weak values is that they are complex, leading
to discussion of the significance of the imaginary part. Here, we show how
one may relate the imaginary part of the momentum weak value to a physically
measurable velocity. For this purpose we consider time averaging, for
example, the mean time it takes the particle to reach the post-selected
position $x$:%
\begin{equation}
\left\langle t\left( x\right) \right\rangle \equiv \int_{0}^{\infty }dtt\rho
\left( t|x\right)
\label{21.1} .
\end{equation}%
This is an experimentally measurable quantity, it implies placing a
\textquotedblleft screen\textquotedblright\ at the position $x$ and then
measuring the time of flight of particles exiting a source and reaching the
screen. The mean time is just $\left\langle t\left( x\right) \right\rangle $%
. We can repeat this measurement at two successive values of $x$ which are
close to each other and in this way also measure how this mean time changes
with the position of the screen. Specifically%
\begin{eqnarray}
&&\noindent \frac{\partial \left\langle t\left( x\right) \right\rangle }{\partial x}=
\int_{0}^{\infty }dtt\frac{\partial }{\partial x}\rho \left( t|x\right)= \notag
\\ &&
=-\frac{1}{N\left( x\right) }\frac{\partial N\left( x\right) }{\partial x}%
\left\langle t\left( x\right) \right\rangle +\frac{1}{N\left( x\right) }%
\int_{0}^{\infty }dtt\frac{\partial \left\vert \langle x|\Psi _{t}\rangle
\right\vert ^{2}}{\partial x},\notag \\
\label{21.2}
\end{eqnarray}
where $N(x)$ has been defined in  {  Eq. \ref{18}.} On the
other hand the imaginary part of the weak value of the momentum as {  seen
from Eq. 21 is:}%
\begin{equation}
\mathrm{Im}\left\langle \hat{p}\left( x;\Psi _{t}\right)
\right\rangle _{w}=-\frac{\hbar }{2\left\vert \langle
x|\Psi _{t}\rangle \right\vert ^{2}}\frac{\partial \left\vert \langle x|\Psi
_{t}\rangle \right\vert ^{2}}{\partial x}
\label{21.3}
\end{equation}%
so that its time-averaged value is%
\begin{eqnarray}
&&\left\langle \mathrm{Im}\left\langle \hat{p}\left( x;\Psi \right)
\right\rangle _{w}\right\rangle \equiv\int_{0}^{\infty }dt\rho \left(
t|x\right) \mathrm{Im}\left\langle \hat{p}\left( x;\Psi _{t}\right)
\right\rangle _{w}\notag  \\
&=&-\frac{\hbar }{2}\int_{0}^{\infty }dt\left[ \frac{\partial \rho \left(
t|x\right) }{\partial x}+\rho \left( t|x\right) \frac{\partial \ln N\left(
x\right) }{\partial x}\right] \notag \\
&=&-\frac{\hbar }{2}\frac{\partial \ln N\left( x\right) }{\partial x}.
\label{21.4}
\end{eqnarray}%
We thus find that%
\begin{eqnarray}
\frac{\partial \left\langle t\left( x\right) \right\rangle }{\partial x}=%
\frac{2}{\hbar }\left\langle \mathrm{Im}\left\langle \hat{p}\left( x;\Psi
\right) \right\rangle _{w}\right\rangle \left\langle t\left( x\right)
\right\rangle \notag  \\-\frac{2}{\hbar }\int_{0}^{\infty }dtt\rho \left( t|x\right)
\mathrm{Im}\left\langle \hat{p}\left( x;\Psi _{t}\right) \right\rangle _{w},
\label{21.5}
\end{eqnarray}%
which shows how the imaginary part of the weak value of the momentum
determines $\frac{\partial \left\langle t\left( x\right) \right\rangle }{%
\partial x}$ and this in turn may be considered as the inverse of a mean
velocity of the particle at the point $x$.

\section{VI. Inferring the real part of the weak value of the momentum}

Instead of measuring the transition path time distribution as defined above,
one may also measure the number of particles per unit time arriving at the
post-selected point $x$ at the time $t$. The experiment one has in mind is
the following. Initially, one prepares particles described by the initial
wavefunction as before. They will escape from the source. The shutter of the
source is opened for a time $Dt$ which is much shorter than the time it
takes them to arrive at the post-selected point $x$. During this time $Dt$
we assume that $N_{i}$ particles came out of the source. This means that
initially, around $t=0$ the number of particles per unit time exiting the
source is $N_{i}/Dt$. Now one post-selects the point $x$ in the asymptotic
products region (to the right of the potential) and measures the number of
particles per unit time crossing this point at the time $t$. This is the
flux of particles at $x$ at time $t$. Different particles will arrive at
different times at $x$ so that one can measure the flux distribution at $x$
at time $t$. In principle, not all particles will be transmitted. The
transmission probability for particles reaching the post-selected point $x$
is by definition the ratio of the number of particles reaching the screen located at
$x$ ($N_f$) to the total number of incident particles coming out of the source located at $x_i$ ($N_i$)
\begin{equation}
T=\frac{N_f}{N_i}\equiv \frac{\int_{0}^{\infty }dt\left\langle \Psi _{t}\left\vert {\hat F}\left(
x\right) \right\vert \Psi _{t}\right\rangle }{\int_{-Dt/2}^{Dt/2}dt\left%
\langle \Psi _{t}\left\vert {\hat F}\left(x_i\right) \right\vert \Psi
_{t}\right\rangle },  \label{23}
\end{equation}%
where ${\hat F}(x) $ is the flux operator defined in Eq. \ref{9}.

The analog of the transition path time distribution is then the normalized
flux time distribution at the post-selected point $x$:
\begin{eqnarray}
&&f\left( t|x\right) =\frac{\left\langle \Psi _{t}\left\vert {\hat{F}}\left(
x\right) \right\vert \Psi _{t}\right\rangle }{T\int_{-Dt/2}^{Dt/2}dt\left%
\langle \Psi _{t}\left\vert {\hat{F}}\left(x_i\right) \right\vert \Psi
_{t}\right\rangle }=  \notag \\
&=& \frac{\left\langle \Psi _{t}\left\vert {\hat{F}}\left( x\right)
\right\vert \Psi _{t}\right\rangle }{\int_{0}^{\infty }dt\left\langle \Psi
_{t}\left\vert {\hat{F}}\left( x\right) \right\vert \Psi _{t}\right\rangle }
\equiv \frac{\left\langle \Psi _{t}\left\vert {\hat{F}}\left( x\right)
\right\vert \Psi _{t}\right\rangle }{N_{f}}  \label{24}
\end{eqnarray}%
and we note that $N_{f\text{ }}$ is independent of $x$ due to the
conservation of flux.

Using the definition of the flux operator as in Eq. \ref{9} and the momentum
operator as in Eq. \ref{20}, the normalized flux time distribution may be
rewritten as:
\begin{equation}
f\left( t|x\right) =\frac{N\left( x\right) }{MN_{f}}\rho \left( t|x\right)
\mathrm{Re}\left[ \frac{\left\langle x\left\vert {\hat{p}}\right\vert \Psi
_{t}\right\rangle }{\left\langle x|\Psi _{t}\right\rangle }\right]
\label{25}
\end{equation}%
and this is identical to the formal result given in
Eq. \ref{12}. In words, the real part of the weak value of the momentum at the
post-selected point $x$ is proportional to the ratio of the flux and density
time distributions. Hence there is also no need to use weak measurement to
obtain the real part of the weak value of the momentum.

\section{VII. Bohmian trajectories and weak momentum value time evolution}

We shall now revisit the role of weak values within Bohmian mechanics in the context of their determination via strong measurements.

We consider a particle with mass $M$ moving under the influence of a potential energy $V(x)$. In Bohmian mechanics the time dependent wavefunction of the particle is represented as:%
\begin{equation} \label{Bohmian}
\left\langle x|\varphi _{t}\right\rangle =\sqrt{r\left( x,t\right) }\exp %
\left[ i\frac{S\left( x,t\right) }{\hbar }\right],
\end{equation}%
where $r\left( x,t\right) $ is a positive function - the density, and $%
S\left( x,t\right) $ is a real valued phase.
It is well known that the time dependent Schr\"odinger equation may be written in terms of the time dependent density and phase as:
\begin{eqnarray}
\frac{\partial S\left( x,t\right) }{\partial t}+\frac{1}{2M}\left[ \frac{%
\partial S\left( x,t\right) }{\partial x}\right] ^{2}+V_{eff}\left(
x,t\right) &=&0 \label{25.1} \\
\frac{\partial r\left( x,t\right) }{\partial t}+\frac{1}{M}\frac{\partial }{%
\partial x}\left[ r\left( x,t\right) \frac{\partial S\left( x,t\right) }{%
\partial x}\right] &=&0,
\label{25.2}
\end{eqnarray}%
where the effective potential is:%
\begin{equation}
V_{eff}\left( x,t\right) =V\left( x\right) -\frac{\hbar ^{2}}{2M\sqrt{%
r\left( x,t\right) }}\left( \frac{d^{2}}{dx^{2}}\sqrt{r\left( x,t\right) }%
\right) .
\label{25.3}
\end{equation}%
In Bohmian mechanics the time-dependent momentum is identified as the spatial derivative of the phase%
\begin{equation}
p_{B}\left( x,t\right) \equiv \frac{\partial S\left( x,t\right) }{\partial x}%
=\mathrm{{Re}\left[ \frac{\left\langle x\left\vert p\right\vert \varphi
_{t}\right\rangle }{\left\langle x|\varphi _{t}\right\rangle }\right]}
\label{25.4}
\end{equation}%
and this connects the real part of the weak value of the momentum with the
Bohmian momentum. Notice though that with this formulation the coordinate $x$
does not vary with time, it is our post-selected point.

One may however ``measure'' the real part of {  the} momentum at different values of the coordinate. Bohmian trajectories are defined by allowing the coordinate to change with time by using the classical equation of motion for its time derivative. One then has the following coupled set of equations%
\begin{eqnarray}
M\frac{dx}{dt} &=&p_{B} \label{25.5}\\
\frac{dp_{B}}{dt} &=&-\frac{dV_{eff}\left( x\right) }{dx} \label{25.6}
\end{eqnarray}%
and these define the Bohmian trajectory $x(t),p_{B}(t)$.

If, however, one keeps the post-selected coordinate $x$ fixed in time one finds that
\begin{eqnarray}
\frac{dp_{B}\left( x,t\right) }{dt} =
\frac{d}{dt}\mathrm{{Re}\left[ \frac{\left\langle x\left\vert p\right\vert \varphi _{t}\right\rangle }{\left\langle x|\varphi _{t}\right\rangle }\right] =}\\
=-\frac{\partial }{\partial x}\left( V_{eff}\left( x,t\right) +\frac{%
p_{B}^{2}\left( x,t\right) }{2M}\right)
\label{25.7}
\end{eqnarray}%
and this differs from the time evolution of the Bohmian momentum. The time evolution of the real part of the weak value of
the momentum at the fixed post-selected state $\vert x\rangle$ is not identical to the time evolution of the momentum of the Bohmian trajectory.

Suppose though that we allow the coordinate to be a function of time, such
that indeed $M\frac{dx}{dt}=p_{B}(x,t)$. Then we have that:%
\begin{eqnarray}
\noindent && \frac{dp_{B}\left( x,t\right) }{dt} = \notag \\ && =\frac{\partial }{\partial t}\mathrm{{%
Re}\left[ \frac{\left\langle x\left\vert p\right\vert \varphi
_{t}\right\rangle }{\left\langle x|\varphi _{t}\right\rangle }\right] +\frac{%
p_{B}\left( x,t\right) }{M}\frac{\partial }{\partial x}{Re}\left[ \frac{%
\left\langle x\left\vert p\right\vert \varphi _{t}\right\rangle }{%
\left\langle x|\varphi _{t}\right\rangle }\right] } \notag \\
&& =-\frac{\partial }{\partial x}V_{eff}\left( x,t\right)
\label{25.8}
\end{eqnarray}%
and we have regained the Bohmian trajectory equation. In this case, the evolution of the coordinate is
not through the propagator. If we define the time dependence of the
momentum using the Heisenberg time evolution operator so that%
\begin{eqnarray}
\noindent && P\left( x,t\right) =\mathrm{{Re}\left[ \frac{\left\langle x\left\vert
p_{t}\right\vert \varphi \right\rangle }{\left\langle x|\varphi
\right\rangle }\right] =} \notag \\ &&=\mathrm{{Re}\left[ \frac{\left\langle x\left\vert \exp
\left( \frac{i}{\hbar }Ht\right) p\exp \left( -\frac{i}{\hbar }Ht\right)
\right\vert \varphi \right\rangle }{\left\langle x|\varphi \right\rangle }%
\right] }
\label{25.9}
\end{eqnarray}%
then:%
\begin{eqnarray}
\frac{dP\left( x,t\right) }{dt} &=&\mathrm{{Re}\left[ \frac{i}{\hbar }\frac{%
\left\langle x\left\vert \exp \left( \frac{i}{\hbar }Ht\right) \left[ H,p%
\right] \exp \left( -\frac{i}{\hbar }Ht\right) \right\vert \varphi
\right\rangle }{\left\langle x|\varphi \right\rangle }\right] }\notag  \\
&=&-\mathrm{{Re}\left[ \frac{\left\langle x\left\vert \exp \left( \frac{i}{%
\hbar }Ht\right) \frac{dV\left( q\right) }{dq}\exp \left( -\frac{i}{\hbar }%
Ht\right) \right\vert \varphi \right\rangle }{\left\langle x|\varphi
\right\rangle }\right] }\notag \\
\label{25.10}
\end{eqnarray}%
which is just the Ehrenfest equation. When considering the transition path time distribution we are ``measuring'' the weak momentum value at a fixed post-selected coordinate $x$ and a fixed time $t$. From the Bohmian point of view the transition path time distribution will then involve contributions from different Bohmian trajectories. However, one does not need to determine them to obtain the distribution.

\subsection{Osmotic velocity and Bohmian potential}

In analogy to the Bohmian momentum associated with the real part of the weak momentum value we may define an ``osmotic'' momentum associated with its imaginary part
\begin{equation}
p_O(x,t) \equiv -\frac{\hbar }{2r\left(
x,t\right) }\frac{\partial r\left( x,t\right) }{\partial x} =\mathrm{{Im}\left[ \frac{\left\langle x\left\vert p\right\vert \varphi
_{t}\right\rangle }{\left\langle x|\varphi _{t}\right\rangle }\right]}.
\label{25.11}
\end{equation}%
The velocity $v_O\equiv p_O/m$ is often called the ``osmotic'' \cite{nelson66,bohm89} or ``diffusive'' velocity \cite{dlp12}, as it is related to changes in the density rather than the phase. Furthermore, the resulting pre-factor $D \equiv i\hbar/2m$ is often interpreted as an imaginary diffusion coefficient within stochastic quantum mechanics \cite{nelson66}.

The kinetic term of the total energy may be defined as $T_B=p_B^2/2M$. Similarly one may define a non-negative internal energy $I_O=p_O^2/2M$ \cite{heifetz15}.  This definition is meaningful because one finds that the mean of the total energy is:
\begin{eqnarray}
\noindent && \langle H \rangle = \int\Psi^{*}(x,t)\left(-\frac{\hbar^2}{2M}\frac{\partial^2}{\partial^2 x} +V(x) \right)\Psi(x,t) dx=\notag  \\ && =\langle T_B + I_O + V \rangle = \text{const.}
\label{25.12}
\end{eqnarray}
is a conserved quantity. By its definition, the internal energy $I_O$ is related to the quantum potential $Q=-\frac{\hbar^2}{2mr}\frac{\partial^2 r}{\partial x^2}$, since $\langle Q \rangle = \langle I \rangle$. The quantum potential in turn affects the dynamics of the Bohmian momentum  $p_B$:
\begin{equation}
\left(\frac{\partial}{\partial t}+p_B\frac{\partial}{\partial x}\right)p_B=-\frac{\partial}{\partial x}\left(Q+V\right).
\label{25.13}
\end{equation}
The imaginary part of the weak momentum value thus reveals the dynamics underlying the Bohmian trajectories which is expressed by the real part. Therefore, both real and imaginary parts of the momentum weak value play important roles in Bohmian mechanics {  and as shown, both can be strongly inferred.}


{ \section{VIII.  Neutron interferometry experiments}}

The purpose of this section is to show the connection between our {  formal } results
and the recent neutron interferometry experiments of Refs. \cite%
{denkmayr17,sponar18}, which also demonstrate how a strong measurement {  may be used to infer }weak values. {  The experiments employed } a combined system and measuring device.
The interferometer creates the neutron ``paths'' whose two possible
``states'' are denoted by $P$. The neutron spin denoted by $S$ is used as a
probe or meter. {  A pre-selected state is prepared as}
\begin{equation}
|\Psi _{i}\rangle =|P_{i}\rangle |S_{i}\rangle,
\label{25.14}
\end{equation}%
where $|P_{i}\rangle$ are the initial path spin states and $|S_{i}\rangle$
the spin states. {  in the experiments the initial spin state was chosen} to be positive in the $x$
direction
\begin{equation}
|S_{i}\rangle =|S_{x};+\rangle,
\label{25.15}
\end{equation}%

The magnetic field is applied in the $z$ direction with field strength given
by $\alpha $. After the scattering event is over, considering only the
interaction Hamiltonian which is linear in the path and spin operators, they
show that the initial pre-selected state changes to
\begin{equation}
|\Psi _{i}\left( \alpha \right) \rangle =\cos \left( \frac{\alpha }{2}%
\right) |P_{i}\rangle |S_{x};+\rangle -i\hat{\sigma}_{z}^{P}\sin \left(
\frac{\alpha }{2}\right) |P_{i}\rangle |S_{x};-\rangle,
\label{25.16}
\end{equation}%
where $\hat{\sigma}_{z}^{P}$ is the path spin operator in the $z$ direction.
The post-selected state in the path direction is denoted as $|P_{f}\rangle $
and the weak value of interest is:%
\begin{equation}
\left\langle \hat{\sigma}_{z}^{P}\right\rangle _{w}=\frac{\left\langle
P_{f}\left\vert \hat{\sigma}_{z}^{P}\right\vert P_{i}\right\rangle }{%
\left\langle P_{f}|P_{i}\right\rangle }.
\label{25.17}
\end{equation}%
The post-selected state of path and probe can take six forms:%
\begin{equation}
|\Psi _{f}\left( j;\pm \right) \rangle =|P_{f}\rangle |S_{j};\pm \rangle,%
\text{ \ \ }j=x,y,z
\label{25.18}
\end{equation}%
that is, the probe may be strongly measured in any of the $x,y,z$
directions, and may point either up or down.

Following the notation as {  in Eqs. \ref{4}-\ref{6},} the density operator
associated with the post-selected state is%
\begin{equation}
\hat{D}\left( j;\pm \right) =|\Psi _{f}\left( j;\pm \right) \rangle \langle
\Psi _{f}\left( j;\pm \right) |,\text{ \ \ }j=x,y,z.
\label{25.19}
\end{equation}%
The ``flux'' operator associated with the density and with the operator
whose weak value is to be determined is:%
\begin{equation}
\hat{F}\left( j;\pm \right) =\frac{1}{2}\left[ \hat{\sigma}_{z}^{P}\hat{D}%
\left( j;\pm \right) +\hat{D}\left( j;\pm \right) \hat{\sigma}_{z}^{P}\right]%
.
\label{25.20}
\end{equation}%
and finally the ``hermitian commutator'' operator takes the form%
\begin{equation}
\hat{C}\left( j;\pm \right) =\frac{i}{2}\left[ \hat{\sigma}_{z}^{P}\hat{D}%
\left( j;\pm \right) -\hat{D}\left( j;\pm \right) \hat{\sigma}_{z}^{P}\right]%
.
\label{25.21}
\end{equation}

The strong value of the density in the $x$ direction with positive spin,
using the ``time'' evolved pre-selected state $|\Psi _{i}\left( \alpha
\right) \rangle $ is found to be after a bit of algebra:
\begin{eqnarray}
\left\langle \Psi _{i}\left( \alpha \right) \left\vert \hat{D}\left(
x;+\right) \right\vert \Psi _{i}\left( \alpha \right) \right\rangle =\cos
^{2}\left( \frac{\alpha }{2}\right) \left\vert \left\langle
P_{i}|P_{f}\right\rangle \right\vert ^{2}\equiv I_{x+} \notag \\
\label{25.22}
\end{eqnarray}%
and this is precisely Eq. 10a of the paper by Sponar \textit{et al.} \cite%
{sponar18}.

The strong value of the flux operator with the probe in the $x$ direction
with positive spin is similarly found to be:%
\begin{eqnarray}
\left\langle \Psi _{i}\left( \alpha \right) \left\vert \hat{F}\left(
x;+\right) \right\vert \Psi _{i}\left( \alpha \right) \right\rangle
=\left\vert \left\langle P_{i}|P_{f}\right\rangle \right\vert ^{2}\cos
^{2}\left( \frac{\alpha }{2}\right) \text{Re}\left\langle \hat{\sigma}%
_{z}^{P}\right\rangle _{w},\notag \\
\label{25.23}
\end{eqnarray}%
and the hermitian commutator operator is%
\begin{eqnarray}
\left\langle \Psi _{i}\left( \alpha \right) \left\vert \hat{C}\left(
x;+\right) \right\vert \Psi _{i}\left( \alpha \right) \right\rangle
=\left\vert \left\langle P_{i}|P_{f}\right\rangle \right\vert ^{2}\cos
^{2}\left( \frac{\alpha }{2}\right) \text{Im}\left\langle \hat{\sigma}%
_{z}^{P}\right\rangle _{w}.\notag \\
\label{25.24}
\end{eqnarray}%
Eqs. {  \ref{5} and \ref{6} }are thus specified to
\begin{equation}
\frac{\left\langle \Psi _{i}\left( \alpha \right) \left\vert \hat{F}\left(
x;+\right) \right\vert \Psi _{i}\left( \alpha \right) \right\rangle }{%
\left\langle \Psi _{i}\left( \alpha \right) \left\vert \hat{D}\left(
x;+\right) \right\vert \Psi _{i}\left( \alpha \right) \right\rangle }=\text{%
Re}\left\langle \hat{\sigma}_{z}^{P}\right\rangle _{w}
\label{25.25}
\end{equation}%
and%
\begin{equation}
\frac{\left\langle \Psi _{i}\left( \alpha \right) \left\vert \hat{C}\left(
x;+\right) \right\vert \Psi _{i}\left( \alpha \right) \right\rangle }{%
\left\langle \Psi _{i}\left( \alpha \right) \left\vert \hat{D}\left(
x;+\right) \right\vert \Psi _{i}\left( \alpha \right) \right\rangle }=\text{%
Im}\left\langle \hat{\sigma}_{z}^{P}\right\rangle _{w}.
\label{25.26}
\end{equation}

The experimental setup made it only possible to measure densities, as given
in Eqs. 10a-10f of Ref. \cite{sponar18}, not fluxes. They extracted the real
and imaginary parts and the absolute value of the weak value from a
combination of the six densities as given in their Eqs. 11a-11c.
Specifically, their Eqs. 10c and 10d are (in their notation):%
\begin{equation}
I_{y+}-I_{y-}=\sin \alpha \left\vert \left\langle P_{i}|P_{f}\right\rangle
\right\vert ^{2}\text{Re}\left\langle \hat{\sigma}_{z}^{P}\right\rangle _{w}
\label{25.27}
\end{equation}%
\begin{equation}
I_{z+}-I_{z-}=\sin \alpha \left\vert \left\langle P_{i}|P_{f}\right\rangle
\right\vert ^{2}\text{Im}\left\langle \hat{\sigma}_{z}^{P}\right\rangle _{w}
\label{25.28}
\end{equation}%
from which we extract:%
\begin{eqnarray}
\left\vert \left\langle P_{i}|P_{f}\right\rangle \right\vert ^{2}\text{Re}%
\left\langle \hat{\sigma}_{z}^{P}\right\rangle _{w}=\frac{I_{y+}-I_{y-}}{%
\sin \alpha }=\notag \\
=\frac{\left\langle \Psi _{i}\left( \alpha \right) \left\vert
\hat{F}\left( x;+\right) \right\vert \Psi _{i}\left( \alpha \right)
\right\rangle }{\cos ^{2}\left( \frac{\alpha }{2}\right) }
\label{25.29}\end{eqnarray}%
and%
\begin{eqnarray}
\left\vert \left\langle P_{i}|P_{f}\right\rangle \right\vert ^{2}\text{Im}%
\left\langle \hat{\sigma}_{z}^{P}\right\rangle _{w}=\frac{I_{z+}-I_{z-}}{%
\sin \alpha }=\notag \\
=\frac{\left\langle \Psi _{i}\left( \alpha \right) \left\vert
\hat{C}\left( x;+\right) \right\vert \Psi _{i}\left( \alpha \right)
\right\rangle }{\cos ^{2}\left( \frac{\alpha }{2}\right) }
\label{25.30}
\end{eqnarray}%
We then have that:%
\begin{equation}
\left\langle \Psi _{i}\left( \alpha \right) \left\vert \hat{F}\left(
x;+\right) \right\vert \Psi _{i}\left( \alpha \right) \right\rangle =\frac{1%
}{2}\cot \left( \frac{\alpha }{2}\right) \left( I_{y+}-I_{y-}\right)
\label{25.31}
\end{equation}%
and
\begin{equation}
\left\langle \Psi _{i}\left( \alpha \right) \left\vert \hat{C}\left(
x;+\right) \right\vert \Psi _{i}\left( \alpha \right) \right\rangle =\frac{1%
}{2}\cot \left( \frac{\alpha }{2}\right) \left( I_{z+}-I_{z-}\right)
\label{25.32}
\end{equation}
so that%
\begin{eqnarray}
&&\noindent \text{Re}\left\langle \hat{\sigma}_{z}^{P}\right\rangle _{w}=\frac{%
\left\langle \Psi _{i}\left( \alpha \right) \left\vert \hat{F}\left(
x;+\right) \right\vert \Psi _{i}\left( \alpha \right) \right\rangle }{%
\left\langle \Psi _{i}\left( \alpha \right) \left\vert \hat{D}\left(
x;+\right) \right\vert \Psi _{i}\left( \alpha \right) \right\rangle }=\notag \\ && = \frac{1%
}{2}\cot \left( \frac{\alpha }{2}\right) \frac{I_{y+}-I_{y-}}{I_{x+}}
\label{25.33}
\end{eqnarray}%
\begin{eqnarray}
&&\noindent \text{Im}\left\langle \hat{\sigma}_{z}^{P}\right\rangle _{w}=\frac{%
\left\langle \Psi _{i}\left( \alpha \right) \left\vert \hat{C}\left(
x;+\right) \right\vert \Psi _{i}\left( \alpha \right) \right\rangle }{%
\left\langle \Psi _{i}\left( \alpha \right) \left\vert \hat{D}\left(
x;+\right) \right\vert \Psi _{i}\left( \alpha \right) \right\rangle }= \notag \\ && = \frac{1
}{2}\cot \left( \frac{\alpha }{2}\right) \frac{I_{z+}-I_{z-}}{I_{x+}}
\label{25.34}
\end{eqnarray}
and these are Eqs. 11a and 11b in Ref. \cite{sponar18}. It thus becomes
evident that the real and imaginary components of the weak spin values which
they inferred are obtained through a strong measurement of the
generalized density, flux and hermitian commutator operators.
\bigskip

\section{IX. Discussion}

At first, the result that weak measurements are not needed to obtain weak values might seem surprising. Part of the motivation for
introducing weak measurements was to reveal information regarding pre- and
post-selected systems without changing them much during the process. On the other hand, strong measurement almost by definition, alters the
system. However, the strong measurement protocol proposed here, allows to accurately infer the weak value of the
unperturbed system because it is executed exactly at the time of
post-selection. In a given run of an experiment, this strong measurement
coincides with the projective measurement used for performing the
post-selection and hence does not disturb the initial or final states of the
system.

Our protocol is not only consistent with recent experiments \cite{denkmayr17,sponar18} employing neutron interferometry, but in fact
generalizes these schemes from discrete operators to any operator. The comparison with neutron interferometry determination of weak spin values
demonstrates the experimental
feasibility of our protocol. The methods presented in Refs. \cite%
{vallone16,denkmayr17,sponar18} indicate also the possible advantage over
the weak measurement technique in terms of precision and accuracy. The
proposed protocol still bears some similarity to the case of weak
measurements, as it does necessitate accumulating enough statistics over a
large ensemble of similarly prepared pre- and post-selected states.

Although appearing ever more frequently in the physics literature, weak
values are still controversial. The question whether they can be strongly
measured or not is still under debate \cite{vaidman17c}, reflecting on earlier discussions
regarding their conceptual meaning and practical significance. The theorem
derived in this paper provides a new approach for strongly inferring the weak value of operators based on time
of arrival measurements. The protocol needs only projective measurements,
thus strengthening the status of weak values as profound
quantities in the quantum mechanical description of pre- and post-selected
systems. The fact that the proposed protocol also accords well with neutron
interferometry experiments \cite{denkmayr17,sponar18}, which showed that
strong measurements of weak values can outperform weak measurements, further
demonstrates the generality of the result and its practical relevance.

Previously, it was shown using the von Neumann measurement scheme that the
imaginary part of the weak value arises from the disturbance due to coupling
with the measuring pointer. This part thus reflects how the initial state is
unitarily disturbed by the measured observable \cite{dressel12}. On the one
hand, Eq. \ref{6}, which depends on the commutator, accords with this view,
but on the other hand, it suggests an alternative way to understand the
imaginary part in a manner which does not require an auxiliary measuring
pointer. Eqs. \ref{5} and \ref{6} show that both real and imaginary parts
of the weak value are physically significant and that both are amenable to
direct, strong inference. The experimental significance of the imaginary
part of the momentum weak {  value was also discussed}.

The importance of the weak value especially of the momentum operator cannot
be overstressed. The real and imaginary parts of the momentum weak value
allow the reconstruction of the wavefunction since they contain the
necessary information regarding the phase and amplitude of the wavefunction,
respectively. Specifically, representing the wavefunction as $\Psi(x,t)=%
\sqrt{\rho(x,t)}\exp\left[iS(x,t)/\hbar\right]$ the phase may be
reconstructed from Eq. \ref{12}
\begin{equation}
S(x,t)=\int\mathrm{Re}\left[ \frac{\left\langle x\left\vert {\hat{p}}%
\right\vert \Psi _{t}\right\rangle }{\left\langle x|\Psi _{t}\right\rangle }%
\right]dx  \label{26}
\end{equation}
and the density from Eq. \ref{13}
\begin{equation}
\rho(x,t)=e^{-\frac{2}{\hbar}\int\mathrm{Im}\left[ \frac{\left\langle
x\left\vert p\right\vert \Psi _{t}\right\rangle }{\left\langle x|\Psi
_{t}\right\rangle }\right]dx}.  \label{27}
\end{equation}
The more general Eqs. \ref{5} and \ref{6} allow in principle to
reconstruct the wavefunction in any other basis.

Weak values have been also used for reconstructing Bohmian trajectories,
since the real part of the weak value of the momentum is identical to the
Bohmian momentum \cite{wiseman07,kocsis11}. The Bohmian approach is also a
somewhat different route towards reconstructing the wavefunction.

To conclude, we have shown in this paper that weak values need not be
considered only in the context of weak measurement, they may be inferred
directly from a strong measurement protocol. These results will hopefully
pave the way for a better understanding of weak values, as well as for
feasible strong measurement based methods for inferring and using them in
practical applications.

\section{Acknowledgements}

We thank Y. Aharonov, J. Dressel, F. Nori, B.E.Y. Svensson and G. Vallone
for their insightful comments on an early version of this manuscript. This
work has been supported by the Canada Research Chairs (CRC) Program and by
grants from the Israel Science Foundation and the Minerva Foundation, Munich.


\end{document}